\documentclass[pra,twocolumn,aps,superscriptaddress,nofootinbib]{revtex4-1}
\usepackage{amsfonts}
\usepackage{mathtools}
\usepackage[svgnames,table]{xcolor}
\usepackage[T1]{fontenc}
\usepackage{txfonts}
\usepackage{graphicx}
\usepackage{lineno}
\usepackage{gensymb}

\usepackage{hyperref}
\hypersetup{
        colorlinks=true,
        citecolor=SteelBlue,
        filecolor=LimeGreen,
        linkcolor=SlateBlue,
        urlcolor=MediumPurple
}

\usepackage{xcite}
\usepackage{dcolumn}
\usepackage{bm}
\usepackage{amsmath}
\usepackage{breqn}
\usepackage[bbgreekl]{mathbbol}
\usepackage{bbm}
\usepackage{mathrsfs}
\usepackage[permil]{overpic}

\makeatletter
\let\cat@comma@active\@empty
\makeatother
\begin{document}
\preprint{APS/123-QED}
\title{Cryogenic optical shadow sensors for future  gravitational wave detectors}
\author{Amit Singh Ubhi}
\affiliation{Institute for Gravitational Wave Astronomy, School of Physics and Astronomy, University of Birmingham, Birmingham B15 2TT, United Kingdom}
\author{John Bryant}
\affiliation{Institute for Gravitational Wave Astronomy, School of Physics and Astronomy, University of Birmingham, Birmingham B15 2TT, United Kingdom}
\author{David Hoyland}
\affiliation{Institute for Gravitational Wave Astronomy, School of Physics and Astronomy, University of Birmingham, Birmingham B15 2TT, United Kingdom}
\author{Denis Martynov}
\affiliation{Institute for Gravitational Wave Astronomy, School of Physics and Astronomy, University of Birmingham, Birmingham B15 2TT, United Kingdom}

\date{\today}

\begin{abstract}
Displacement sensors have a variety of applications within gravitational wave detectors. The seismic isolation chain of the LIGO core optics utilises optical shadow sensors for their stabilisation. Future upgrades, such as LIGO Voyager, plan to operate at cryogenic temperatures to reduce their thermal noise and will require cryogenic displacement sensors. We present the results of simulations and experimental tests of the shadow sensors embedded in the Birmingham Optical Sensors and Electromagnetic Motors (BOSEMs). We determine that the devices can reliably viability operate at 100\,K. We also show that the performance of the BOSEM sensors improves at cryogenic temperatures.
\end{abstract}

\maketitle

\section{\label{sec:intro}Introduction}

Terrestrial gravitational wave detectors such as Advanced LIGO~\cite{Aasi2015} have detected a number of gravitational wave sources since 2015~\cite{Abbott2016, Abbott2019GWTC1, Abbott2021GWTC2}, varying from compact binary sources such as binary black hole mergers~\cite{Collaboration2016, Collaboration2017,Abbott2020a}, binary neutron star inspirals~\cite{Collaboration2017a}, and potential black hole neutron star mergers~\cite{Abbott2020}. Within their most sensitive band, the aLIGO detectors are limited by quantum shot noise and thermal noises of the main optics~\cite{Martynov_QUCORR_2017, Buikema2020}. To combat the thermal noise, the KAGRA~\cite{Somiya2012, Aso2013} detector cools its core optics. Future upgrades and detectors such as LIGO Voyager~\cite{Adhikari_2020}, the Einstein Telescope~\cite{Maggiore_2020}, and Cosmic Explorer~\cite{Abbott2017,Reitze2019} also plan to operate at cryogenic temperatures to reduce their thermal noise~\cite{Harry_Thermal_2012, Gras_CTN_2017}. In the case of LIGO Voyager, the core optics will have a silicon substrate. The material exhibits good mechanical properties at cryogenic temperatures and also has a negligible thermal expansion coefficient at 123\,K~\cite{Adhikari_2020}.

Similar to the current gravitational wave detectors, future observatories will require a comparable or even more sophisticated seismic isolation system~\cite{Plissi2000, Braccini__2002, BEKER20121389}. The design of the isolation systems is different in LIGO, Virgo, and KAGRA. However, the detectors share the same isolation principle: suspend the test masses using a multi-stage pendulum suspension to passively filter ground vibrations above the suspension resonances. The resonances are damped using Optical Sensors and Electromagnetic Motors (OSEMs)~\cite{Strain2012, Akutsu_2020}. The same devices monitor the position of the suspension chain and provide actuation on its stages.

In this paper, we study the Birmingham design of OSEMs (BOSEMs) at cryogenic temperatures to verify their suitability for the cryogenic detectors. The devices as well as another design known as Advanced LIGO OSEMs currently sense and actuate on the LIGO suspensions. BOSEMs are perfect candidates for the next generation observatories because the devices have proven their reliability, large linear range, and ultra high vacuum compatibility. The key question that we address in the paper is whether BOSEMs can operate at cryogenic temperatures?
In this paper, we find that the answer to the question is positive. However, the behaviour of optical sensors at cryogenic temperatures is slightly different from their operation at room temperature.
We study the discrepancies between the emission of the LED, the photodetector current, and the inherent noise of the device at various temperatures. We also investigate the quantum efficiency of BOSEM LEDs. The efficiency improves at cryogenic temperatures and reduces the overall noise of the sensors.
We discuss the potential concerns of cooling down these devices, analyse their sensitivity and propose changes to the electronic design to optimise them for cryogenic usage.

\section{\label{sec:CAD}Finite-element analysis}

\begin{figure}[!htp]
    \centering
    \includegraphics[width = \columnwidth]{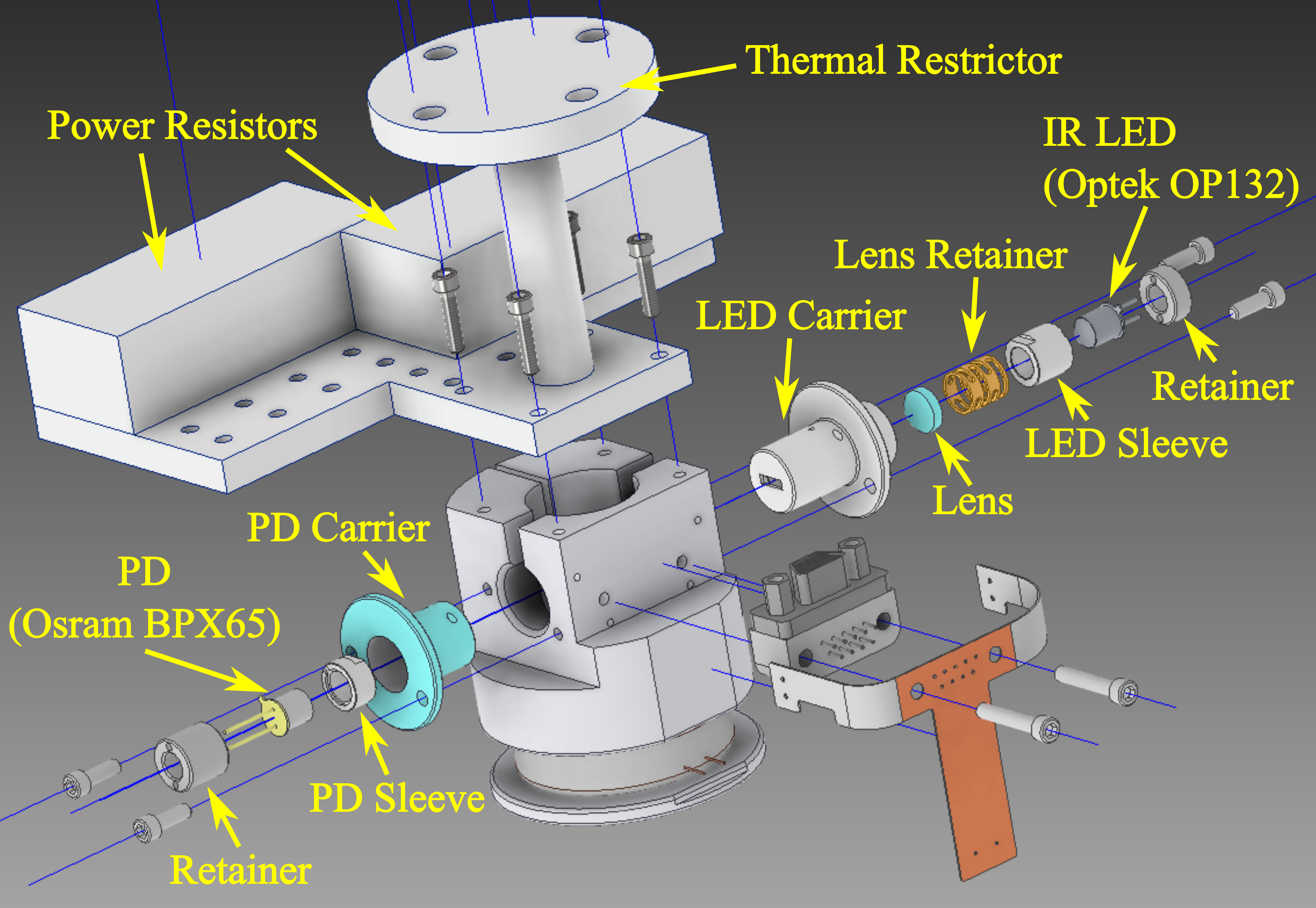}    
    \caption{Exploded view of BOSEM setup.}
    \label{fig:exploded}
\end{figure}

\begin{figure}[!htb]
    \centering
    \includegraphics[width=.9\linewidth]{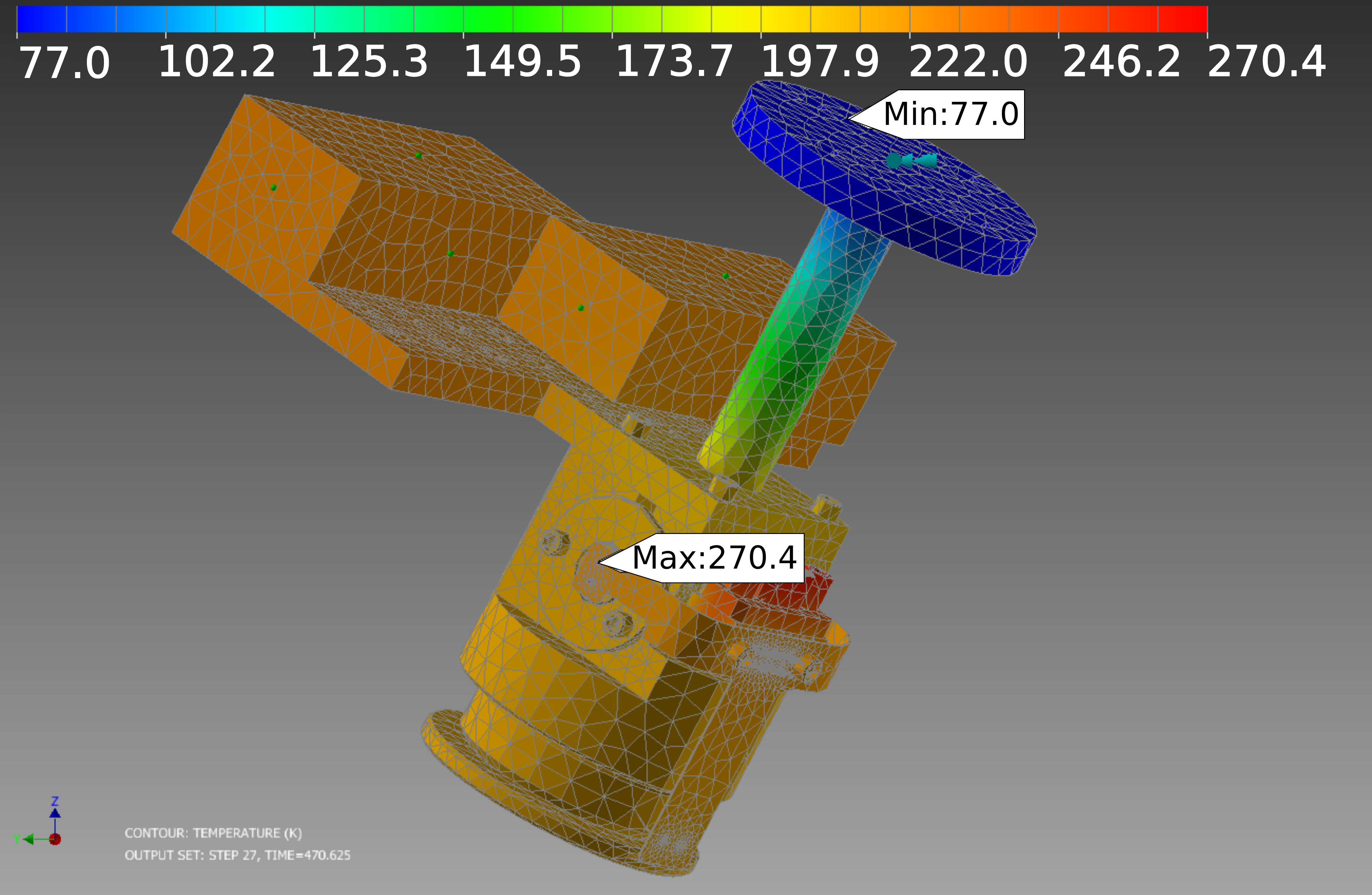}
    \caption{Temperature (K) map of the BOSEM during the cool down simulation. This time step was chosen as it corresponded to maximum stresses in the BOSEM due to temperature variations.}
    \label{fig:temp-map}
\end{figure}
  \begin{figure}[!htb]
    \centering
    \includegraphics[width=.9\linewidth]{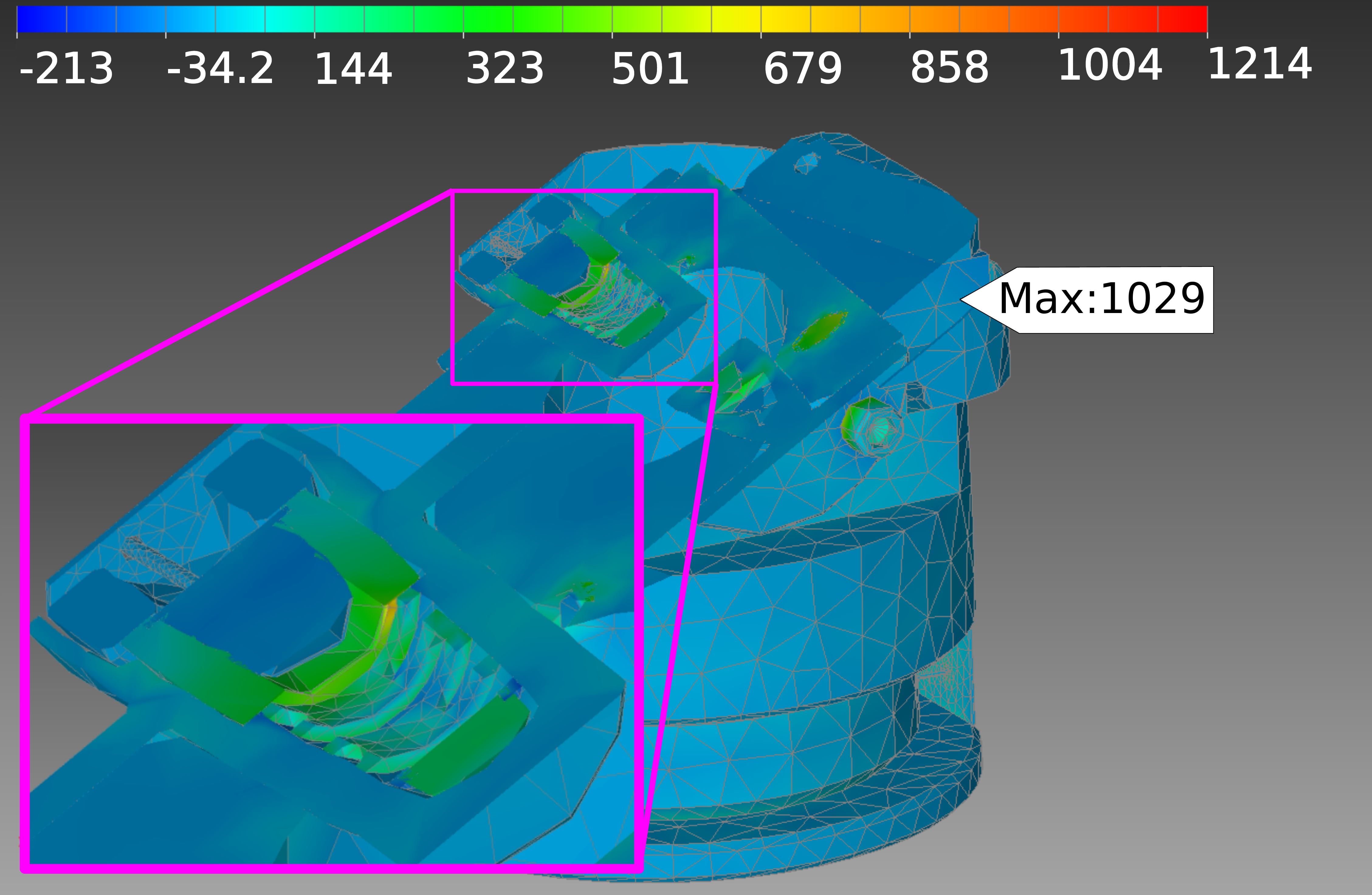}
    \caption{Stress (MPa) map from temperature map in Fig.~\ref{fig:temp-map}. The highlighted and zoomed section in magenta focuses on a cross section of the LED carrier and its components.}
    \label{fig:stress-map}
\end{figure}

Initially, we simulated the cool down procedure of our experimental setup. The setup consists of two main parts: a thermal restrictor and a BOSEM. The restrictor was machined from aluminium and connects the BOSEM to a liquid nitrogen reservoir. The restrictor design was optimised for temperature tuning of the BOSEM with two power resistors located on the restrictor plate as shown in Fig~\ref{fig:exploded}. The figure also shows an exploded view of the BOSEM. The shadow sensing is achieved by an LED and a photodetector (PD): a measured target blocks part of the LED light to change the PD signal~\cite{Carbone2012}. Both the LED and PD are embedded in the separate assemblies consisting of cylindrical parts for stability and electrical insulation.

The photodetector and LED assemblies are the main concerns during the cool down process because the elements are manufactured from materials with different indices of thermal expansion. Therefore, temperature gradients during the cool down process and uneven contraction of the assembly elements can cause stresses in the device. In particular, the LED and photodetector sleeves are made of MACOR for their electrical insulation while their carriers are made of aluminium to simplify the production process. During cool down, the aluminium carriers would contract more than MACOR, applying a compressive load on the sleeves which may potentially lead to fractures or failure of the component. However, the tolerances of the components allow for the slight excess movement of the sleeve within the carrier. The carrier's inner diameter is, $d_{\rm carrier} = 7.290^{+0.015}_{-0.000}\, \rm mm$, 
with the sleeve having outer diameter, $d_{\rm sleeve} = 7.239^{+0.000}_{-0.036}\, \rm mm$. 
Assuming a constant coefficient of linear expansion over the temperature range during cooling, the inner diameter of the carrier is always larger than the the maximum diameter of the sleeve which should result in no compressive stress.

We conducted the analyses of the cool down process with the Autodesk Inventor Nastran add-on~\cite{autodesk_2021}. We achieved a transient heat transfer model with the constraint that the top of the thermal restrictor is held at 77\,K. For the simulation and cool down of the device, two power resistors were used for temperature control. During the simulation they were set with a constant output of 0.2\,W of power. The result of the simulation was a temperature map, shown in Fig.~\ref{fig:temp-map}. We then utilised the results of the heat transfer simulation to calculate the stress on the system due to the temperature gradients. Fig.~\ref{fig:stress-map} is a cross sectional cut of the BOSEM, the highlighted section shows the stress on the MACOR sleeve within the LED carrier. We found that a significant stress comparable to the breaking stress occurs at the interface between the lens retainer and MACOR sleeve. However, the natural length of the lens retainer should reduce when cooled, therefore reducing the axial compressive stress applied against the sleeve. The maximum indicated stress shown in Fig.~\ref{fig:stress-map} occurs between a screw and the body of the BOSEM. Simulation results confirmed the areas of concern during cooling, and experimental tests were needed to verify the performance of the device at cryogenic temperatures.

\section{\label{sec:experiment}Experimental results}

\begin{figure}[!htp]
    \centering
    \includegraphics[width = \columnwidth]{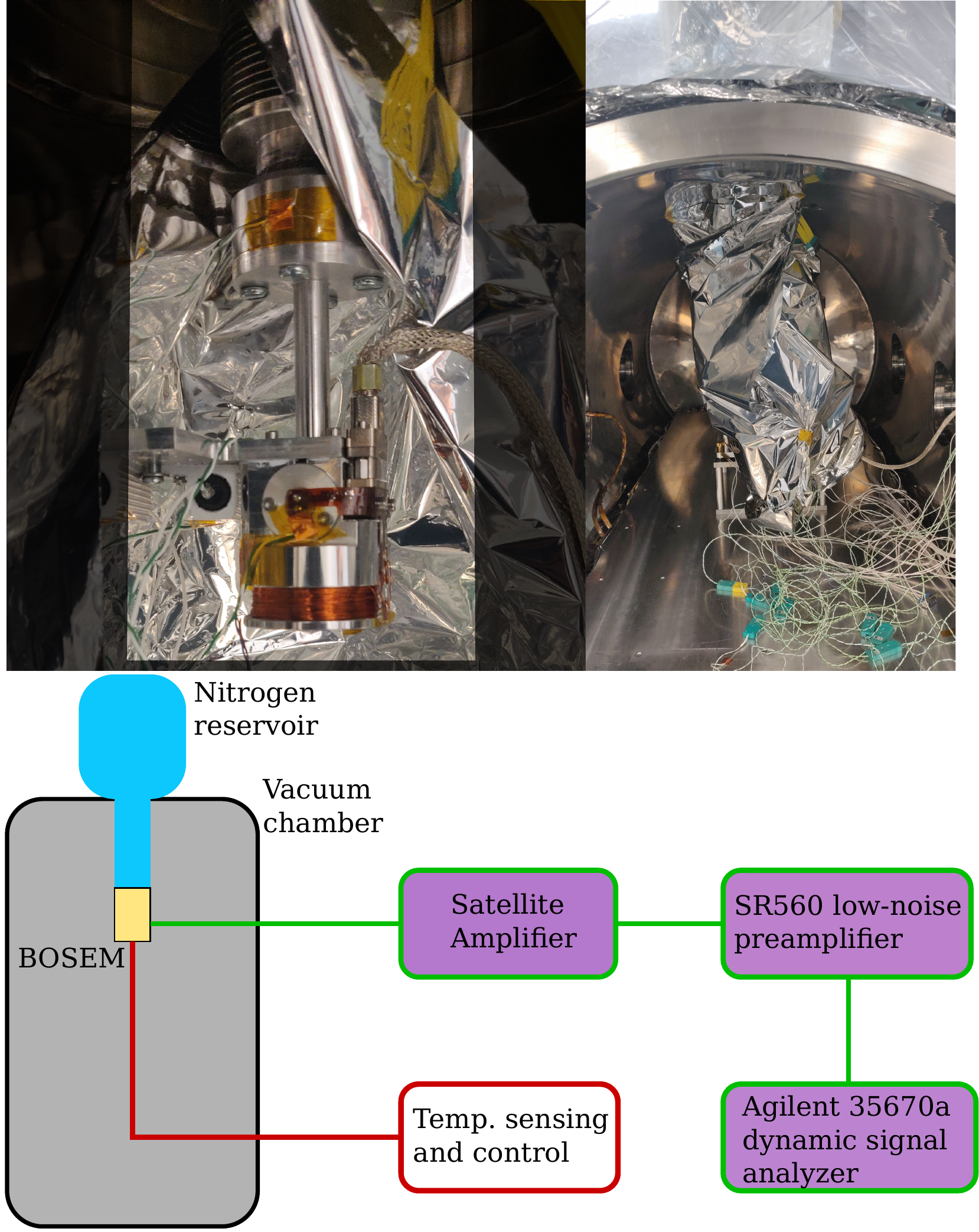}
    \caption{\textbf{Top left}: Photo of the BOSEM attached inside the vacuum chamber. Above the BOSEM is the thermal restrictor with power resistors connected. Thermocouples are attached to measure its temperature. \textbf{Top right}: Image of the BOSEM setup inside the chamber. The set up is encased in mylar to reduce radiative heating from the tank. \textbf{Bottom} The  Schematic of the experimental set up. The BOSEM satellite amplifier output was converted to a single ended signal, AC coupled at 30\,mHz, and high passed at 1\,kHz using an SR560 low-noise preamplifier~\cite{sr560}- its output was measured using the Agilent 35670a dynamic signal analyser~\cite{35670a}. Multiple thermocouples were used to measure the BOSEM's temperature, and power resistors were used for temperature stabilisation.}
    \label{fig:setup}
\end{figure}

The shadow sensors have a linear range of $d = 0.7$\,mm~\cite{Carbone2012} over which they are nominally operated and is used as an ostensible value to calibrate the devices. The measured displacement corresponds to the changes of the current from the photodetector. This current is then converted to voltage via a satellite amplifier~\cite{Carbone2012} as shown in Fig.~\ref{fig:setup}. The observed voltage is then converted to the displacement using the equation

\begin{equation}
    \centering
    K = \frac{V_{\rm max}}{d},
    \label{eq:conversion factor}
\end{equation}
where $V_{\rm max}$ is the maximum voltage on the output of the satellite amplifier. Typical values of the calibration coefficient $K$ are 20-25\,kV/m dependent on the parameters of LEDs and photodetectors. In our setup, we measured $K=20.371$\,kV/m.

The amplifiers signal is usually fed to the LIGOs control and design system for data processing and requires extra whitening of the signal to overcome the noise of the analog-to-digital converters. In our case, the signal was amplified with SR560 and then measured with an AC-coupled Agilent 35670a dynamic signal analyzer with low self-noise. In this paper, we tested the shadow sensors without a measured target. Instead, the measurements were taken with the LED either switched on or off, resulting in maximal or zero light from the LED incident on the PD. A baseline measurement was taken in vacuum at $2\times10^{-4}\,\rm mbar$ to reduce the thermal conductivity between the BOSEM with the environment.

\begin{figure}[!htp]
        \centering
        \includegraphics[width = \columnwidth]{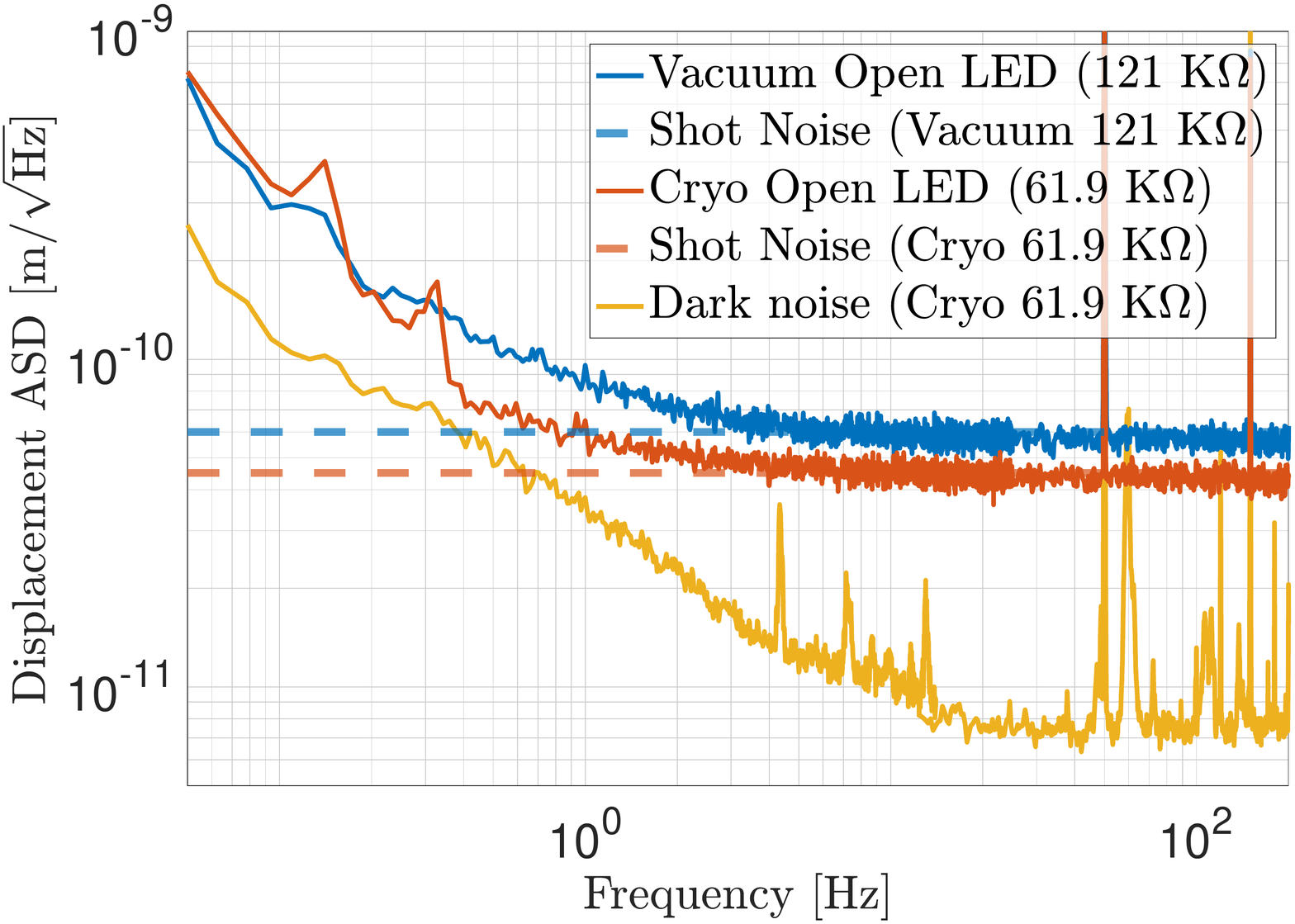}
        \caption{Comparisons of the measurements made in vacuum and below 123\,K. The high frequency shot noise improvements come from the increased PD current. The measurements contain low frequency peaks which occur due to the temperature stabilisation controller. The $61.9\,\rm k\Omega$ transimpenance at room temperature coincided with the blue curve at room temperature.}
        \label{fig:comparison}
    \end{figure}

Above 10\,Hz, the shadow sensors are limited by their photocurrent shot noise spectrum,

\begin{equation}
    \centering
    \sigma_{\rm shot}^2 = 2 e I_{\rm DC}.
    \label{eq:shot-noise}
\end{equation}

where $\sigma_{\rm shot}^2$ is the flat white power spectral density of the current, $e$ is the charge of an electron, and $I_{\rm DC}$ is the DC current~\cite{Bramsiepe2018}. We found the shot noise sensitivity to be $6\times10^{-11} \rm m/\sqrt{\rm Hz}$ above 5\,Hz as shown in Fig.~\ref{fig:comparison}. We note that the value of the shot noise depends on the position on the measured target in the shadow sensor because the $I_{\rm DC}$ changes proportional to the target motion while calibration coefficient $K$ stays the same.




    \begin{figure}[!htp]
        \centering
        \includegraphics[width = \columnwidth]{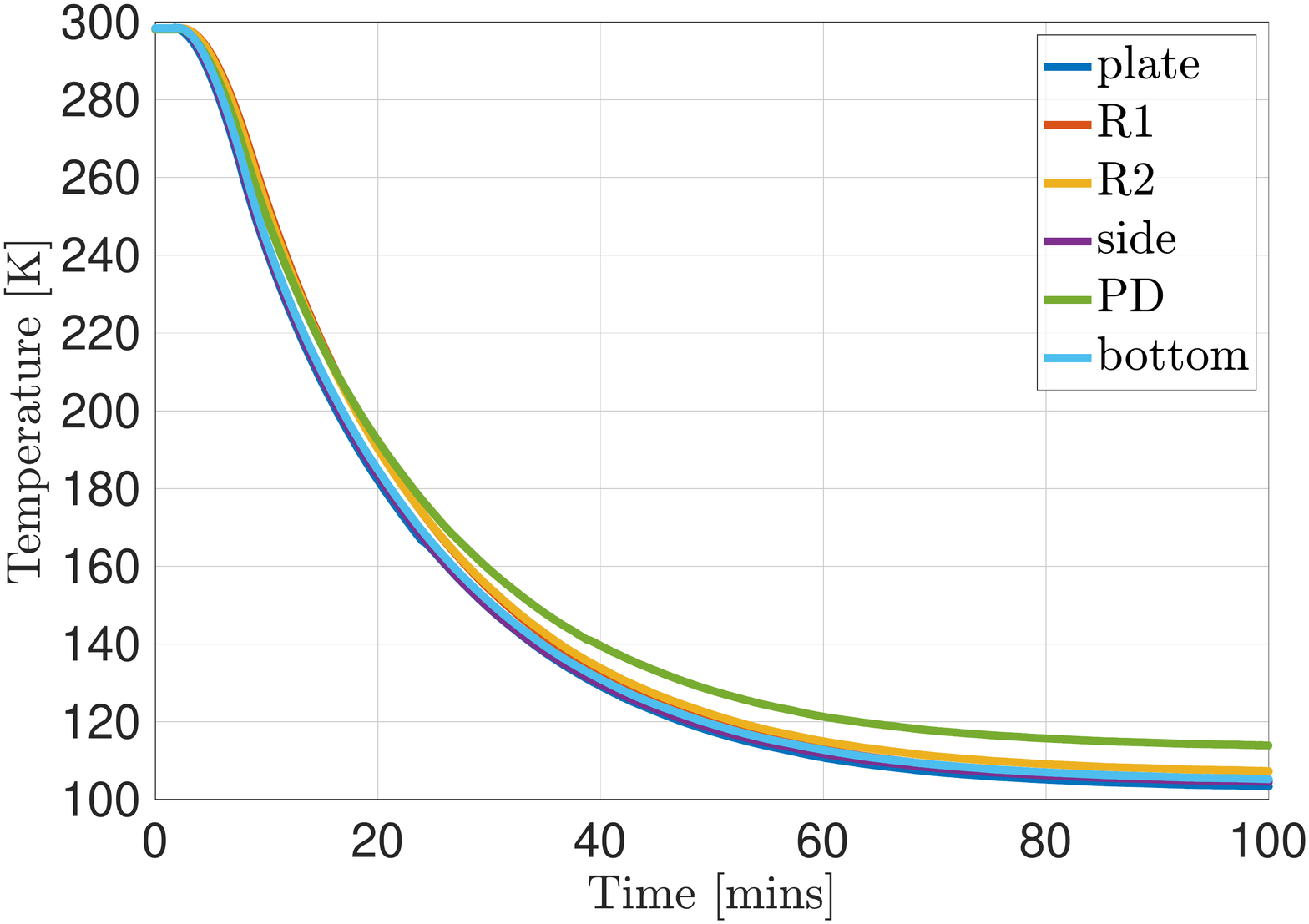}
        \caption{Temperatures of BOSEM during cool down. the blue curve represents the thermal restrictor plate, and red and yellow were the power resistors. Purple was the side of the BOSEM housing. The PD outer casing is green, and the bottom of the BOSEM is in cyan. The inital cool down during the first 20 minutes sees the BOSEM change in temperature by over 100\degree\,C.}
        \label{fig:temperature}
    \end{figure}

We thermocycled the BOSEM multiple times to verify their reliability during and after the cool down process. A number of thermocouples were attached to various parts of the BOSEM to monitor its temperature as shown in Fig.~\ref{fig:temperature}. During testing the PD temperature was used in feedback, and was stabilised to a temperature of 117.7\,K. A temperature below 123\,K was used to ensure that any discrepancy in the thermocouple reading at low temperature was accounted for, and that the BOSEM was cycled past its intended operating temperature which could be the case in future operation. 

    \begin{figure}[!htp]
        \centering
        \includegraphics[width = \columnwidth]{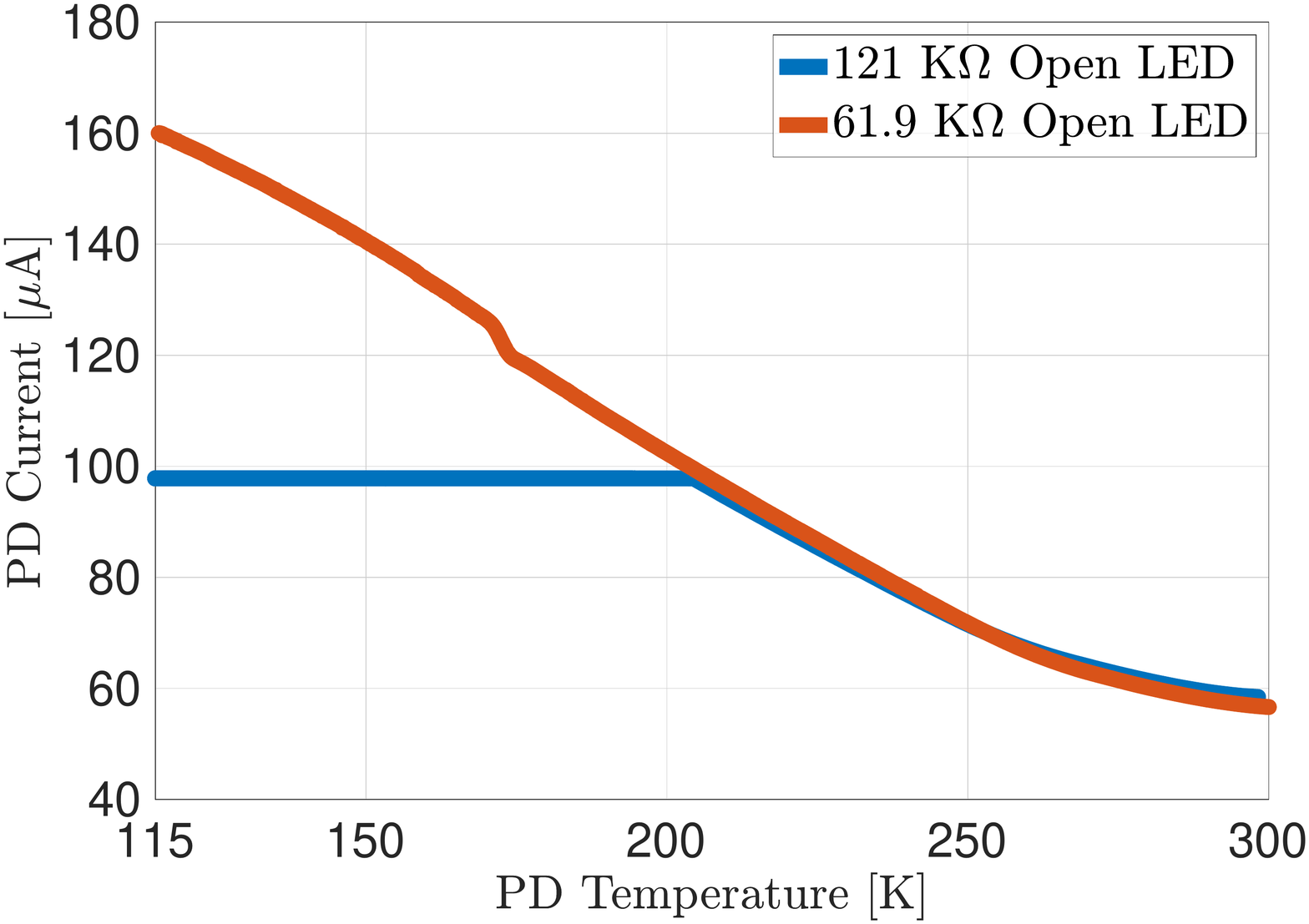}
        \caption{Comparison of the PD current during cool down. The $61.9\,\rm k\Omega$ transimpedance did not cause saturation of the satellite amplifier output unlike the $121\,\rm k\Omega$ transimpedance which saturated below 205\,K.}
        \label{fig:current}
    \end{figure}
    
The response of the device is highly sensitive to temperature, during cool down the PD current increases as shown in Fig.~\ref{fig:current}. A typical satellite amplifier box contains a $121\,\rm k\Omega$ transimpedance amplifier, resulting in the output saturating for a $100\,\mu$A PD current input. Projection of the PD current enabled us to predict the correct amplifier gain to use such that the satellite box output does saturate. The transimpedance was reduced to $61.9\,\rm k\Omega$ while the calibration factor $K$ increased to 27.635\,\rm kV/m at 117.7\,K. Noise improvements from cooling occur due to the improved quantum efficiency of the LED and photodetector, which results in a stronger signal. Overall, the shot noise of the device improves at cryogenic temperatures to $4.5\times 10^{-11}\rm m/\sqrt{\rm Hz}$ as shown in Fig.~\ref{fig:comparison}.

The total efficiency of the optical system is defined according to the equation
\begin{equation}
    \eta_{\rm tot} = \eta_{\rm LED} \eta_{\rm loss} \eta_{\rm PD} =  \frac{I_{\rm PD}}{I_{\rm LED}},
\end{equation}
where $\eta_{\rm LED}$ and $\eta_{\rm PD}$ are the quantum efficiencies of the LED and photodetector and $\eta_{\rm loss}$ is the amount of optical power lost between the LED and PD.
At room temperature, $\eta_{\rm LED} \approx 0.058$ and $\eta_{\rm PD} \approx 0.74$ according to their specifications and $\eta_{\rm tot} = 60\,\mu {\rm A} / 35\,{\rm mA} = 1.7\times 10^{-3}$. Therefore, we find $\eta_{\rm loss} \approx 0.04$. The loss is intentionally introduced by the slit inside the LED carrier (see Fig.~\ref{fig:exploded}) to collimate the beam.

At 117\,K, we measure $\eta_{\rm tot} = 160\,\mu {\rm A} / 35\,{\rm mA} = 4.6\times 10^{-3}$. Since our finite element modelling shows no significant deformation of the optical system at cryogenic temperatures as discussed in Sec.~\ref{sec:CAD}, we expect that the optical loss $\eta_{\rm loss}$ is temperature independent. Therefore, the improvement in $\eta_{\rm tot}$ comes from photodetector and LED efficiencies. However, since $\eta_{\rm PD}$ is already 0.75 at room temperature and the LED central wavelength does not significantly change during the cool down (the spectral shift is 0.3\,nm/K), we conclude that the photodetector cannot be responsible for the increase of $\eta_{\rm tot}$ by the observed factor of 2.7.

The LED efficiency can be written as the product of the internal quantum efficiency of gallium arsenide $\eta_{\rm in} \approx 0.9$ and the external escape efficiency $\eta_{\rm ex}$. The escape efficiency is related to the angle of total internal reflection $\theta$ between the LED active layer and vacuum~\cite{HUI202077}. We note that the formalism in~\cite{HUI202077} neglects the reflection of light on the interface between the LED substrate and its active layer. Therefore, this can be only used as a guidance to understand the cryogenic performance of our LED. As reported in~\cite{McCaulley_1994}, the index of refraction of gallium arsenide reduces at cryogenic temperatures with $dn/dT \approx 10^{-4}\,K^{-1}$ at room temperature. Therefore, we expect that $\eta_{\rm ex}$ improves due to the increase of the angle of total internal incidences inside the LED. We determined $\eta_{\rm LED}$ improved to a value of 0.156 when cooled to 117\,K.





\section{\label{sec:conclusion}Conclusion}

We have demonstrated reliable cryogenic operation of the BOSEMs. The devices have already enabled the control and stabilisation of the LIGO suspensions at room temperature and now have the potential to continue their operation in future gravitational wave detectors. In this paper we have simulated and measured BOSEMs at temperatures below 123\,K. We found that the currently used satellite amplifiers saturate below 205\,K (Fig.~\ref{fig:current}) due to the enhancement of the LED quantum efficiency at cryogenic temperatures. We have reduced the transimpedance from $121\,\rm k\Omega$ down to $61.9\,\rm k\Omega$ to avoid the saturation at cryogenic temperatures. 

Maximising the output of the satellite amplifier also improves the shot noise limited sensitivity of the shadow sensors. We achieved the shot noise level of $4.5\times 10^{-11}\rm m/\sqrt{\rm Hz}$ with a fully open photodetector compared to the room temperature sensitivity of $6\times 10^{-11}\rm m/\sqrt{\rm Hz}$ in the same optical configuration. The improvement came from increased LED quantum efficiency from 0.058 at room temperature up to 0.156 at 117\,K.
The spectrum below 5\,Hz was limited by the LED current drive and temperature stabilisation feedback loop which caused peaks in the spectrum (Fig.~\ref{fig:comparison}).

We also found that the first derivative of the LED current over temperature is significantly larger below 200\,K compared to the 300\,K case. Therefore, temperature fluctuations of BOSEMs should be suppressed in the future gravitational wave detectors for the best sensing noise of the shadow sensors. The cryogenic system of future gravitational wave detectors should also take into account heating from BOSEM LEDs. Even though the quantum efficiency of the LED improves at cryogenic temperatures, approximately 84\% 
of power is still radiated as heat around 100\,K. We estimate that each shadow sensor dissipates $61\,\rm mW$ with a current drive of $35\,\rm mA$.

\section*{Acknowledgements}
We thank members of the LIGO Voyager and SWG groups for useful discussions and Koji Arai for his valuable internal review.
The authors acknowledge the support of the Institute for Gravitational Wave Astronomy at the University of Birmingham, STFC 2018 Equipment Call ST/S002154/1, STFC 'Astrophysics at the University of Birmingham' grant ST/S000305/1. A.S.U. and J.S. are supported by STFC studentships 2117289 and 2116965.

\section*{Bibliography}
\bibliography{bosem.bib}

\end{document}